\documentclass[twocolumn, a4paper, times, trackchanges]{aastex61}

\usepackage[english]{babel}
\usepackage{amsmath}
\usepackage{graphicx}
\usepackage{natbib}

\begin{document}
\title{Energetic neutral atoms from the heliosheath as an additional population of neutral hydrogen in the inner heliosphere}

\correspondingauthor{M. Bzowski}
\email{bzowski@cbk.waw.pl}

\author{M. Bzowski}
\affiliation{Space Research Centre, Polish Academy of Sciences (CBK PAN),\\
Bartycka 18A, 00-716 Warsaw, Poland}

\author{A. Galli}
\affiliation{Physics Institute, University of Bern,\\
Bern, 3012, Switzerland}

\begin{abstract}
Interstellar neutral hydrogen (ISN H) gas penetrates freely the heliopause. Inside the inner heliosheath, the charge-exchange interaction of this gas with the shocked solar wind and pickup ions creates energetic neutral atoms (ENAs). ISN H is strongly depleted inside the termination shock but a fraction reaches the Earth's orbit. In these regions of the heliosphere, ISN H is the source population for interstellar pickup ions and for the heliospheric backscatter glow. The Globally Distributed Flux (GDF) of ENAs created in the inner heliosheath has been sampled directly by Interstellar Boundary Explorer (IBEX). Based on these measurements, we calculate the density of the GDF ENA population at the Earth's orbit. We find that this number density is between $10^{-4}$ and $10^{-3}$~cm~$^{-3}$, i.e., comparable in magnitude to the number density of ISN H in the downwind portion of the Earth's orbit. Half of this atom population have energies less than $\sim 80$~eV. This GDF population of neutral hydrogen is likely to provide a significant contribution to the intensity of heliospheric glow in the downwind hemisphere, may be the source of the inner source of hydrogen pickup ions, and may be responsible for the excess of production of pickup ions found in the analysis of magnetic wave events induced by the proton pickup process in the downwind region at 1~au from the Sun.
\end{abstract}

\section{Introduction}
\label{sec:intro}
The Sun is moving through a cloud of partly ionized, warm, magnetized interstellar matter. The interaction between this gas and the hypersonic solar wind creates the heliosphere \citep{axford:72}. While the ionized component of the Local Interstellar Medium (LISM) is deflected and flows past the heliopause, the neutral component of the LISM, composed mostly of hydrogen \citep[about 0.2~atoms~cm$^{-3}$, ][]{bzowski_etal:09a} and helium \citep[0.015~atoms~cm$^{-3}$, ][]{gloeckler_etal:04a}, is flowing into the heliosphere. Inside the heliosphere, this component is subjected to strong ionization losses and solar resonant radiation pressure, but still it is able to reach the Earth's orbit, albeit heavily depleted \citep{blum_fahr:70a}. The density distribution of interstellar neutral hydrogen (ISN H) along the Earth's orbit has a characteristic pattern, with a deep minimum at the downwind side (on the order of 10$^{-5}$~atoms~cm$^{-3}$) and a maximum at the upwind side \citep[on the order of 10$^{-3}$~atoms~cm$^{-3}$, ][]{thomas:78}. This density is strongly modulated during the solar cycle, with the downwind/upwind density ratio varying by more than an order of magnitude \citep{bzowski_rucinski:95b}. The thermal spread of ISN H at 1~au is strongly anisotropic, but its magnitude is on the order of 10~km~s$^{-1}$, which corresponds to a temperature of $10^4$~K \citep{bzowski_etal:97}.

Direct-sampling measurements of neutral H at 1~au by the Interstellar Boundary Explorer \citep[IBEX,][]{mccomas_etal:09a} showed the presence of ISN H \citep{saul_etal:12a}, strongly evolving during the solar cycle \citep{saul_etal:13a, galli_etal:18a}, but also discovered  the globally distributed flux (GDF) of H atoms \citep{mccomas_etal:09c} at all energies from the upper boundary of IBEX sensitivity at $\sim 6$~keV \citep{funsten_etal:09b} down to $\sim 10$~eV  \citep{fuselier_etal:12a}, as qualitatively predicted by theoretical models \citep{gruntman:97}. As summarized by \citet{mccomas_etal:11b, mccomas_etal:14a}, most likely, the source of these latter atoms are complex charge-exchange reactions between the ambient ISN H atoms and protons from the  plasma inside the heliosphere and beyond the heliopause. The resulting energetic neutral atoms (ENAs) inherit the kinematic properties of the parent protons and due to the lack of electromagnetic interactions run away from their birth sites at distances of several hundred au in all directions, including those sunward. 

In this paper, we focus on the lowest-energy atoms from GDF, observed by the IBEX-Lo instrument \citep{fuselier_etal:09b}. Based on the measurements from 2009.0 to 2012.5, carefully processed by \citet{galli_etal:16a}, we show that by number, the largest contribution to the ENA population at 1~au is given by atoms with lowest energies. Effectively, these atoms make an additional population of neutral H at 1~au, up to now neglected in the analyzes of the creation of PUI fluxes and the backscatter glow in the inner heliosphere. The purpose of this paper is to alert the researcher community to the presence of this population and estimate its number density at 1~au directly based on IBEX observations, as well as to point out some consequences of the existence of this population. A more in-depth analysis of the variation of the GDF gas with time and with the distance from the Sun requires making complex simulations and therefore is postponed to future papers.

\section{Data}
\label{sec:data}
\begin{figure*}
\epsscale{1.10}
\plottwo{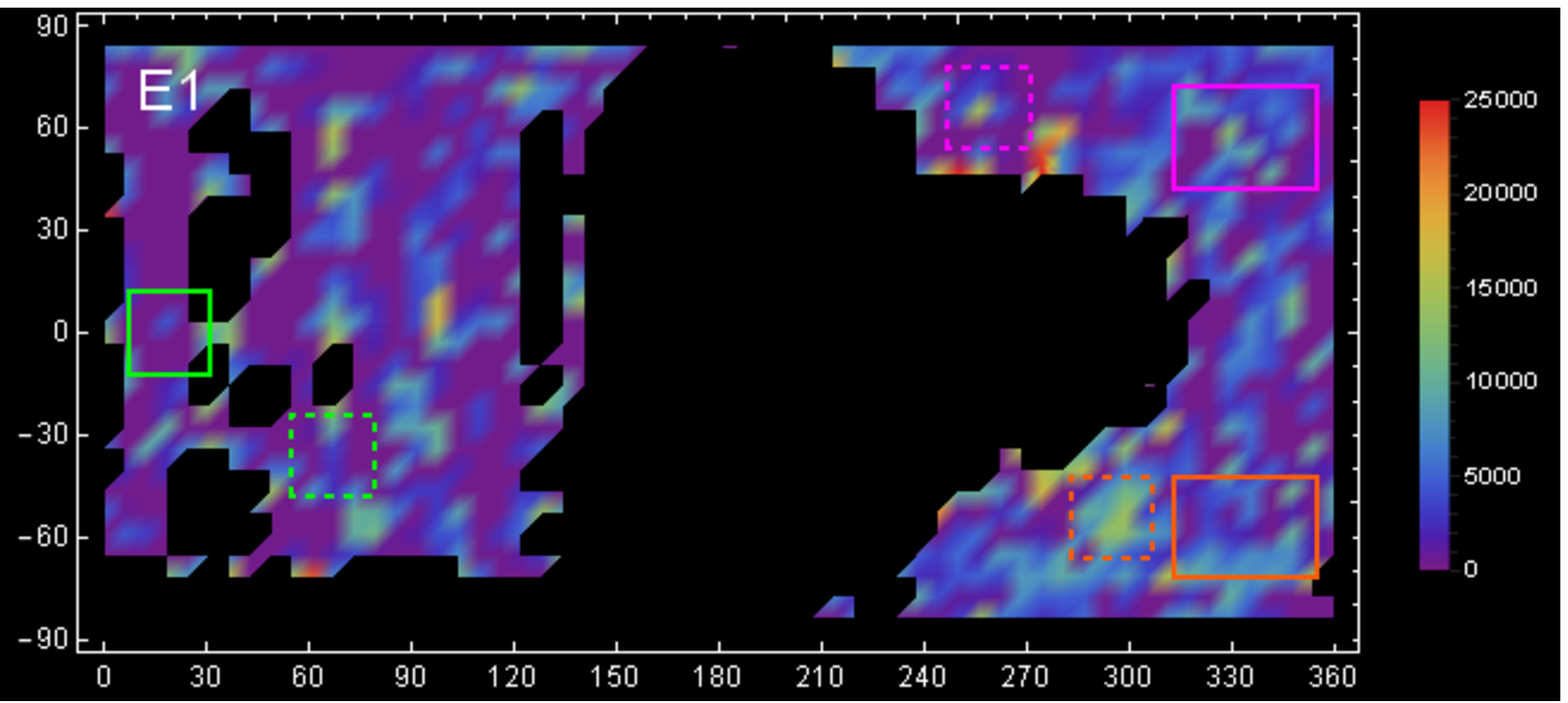}{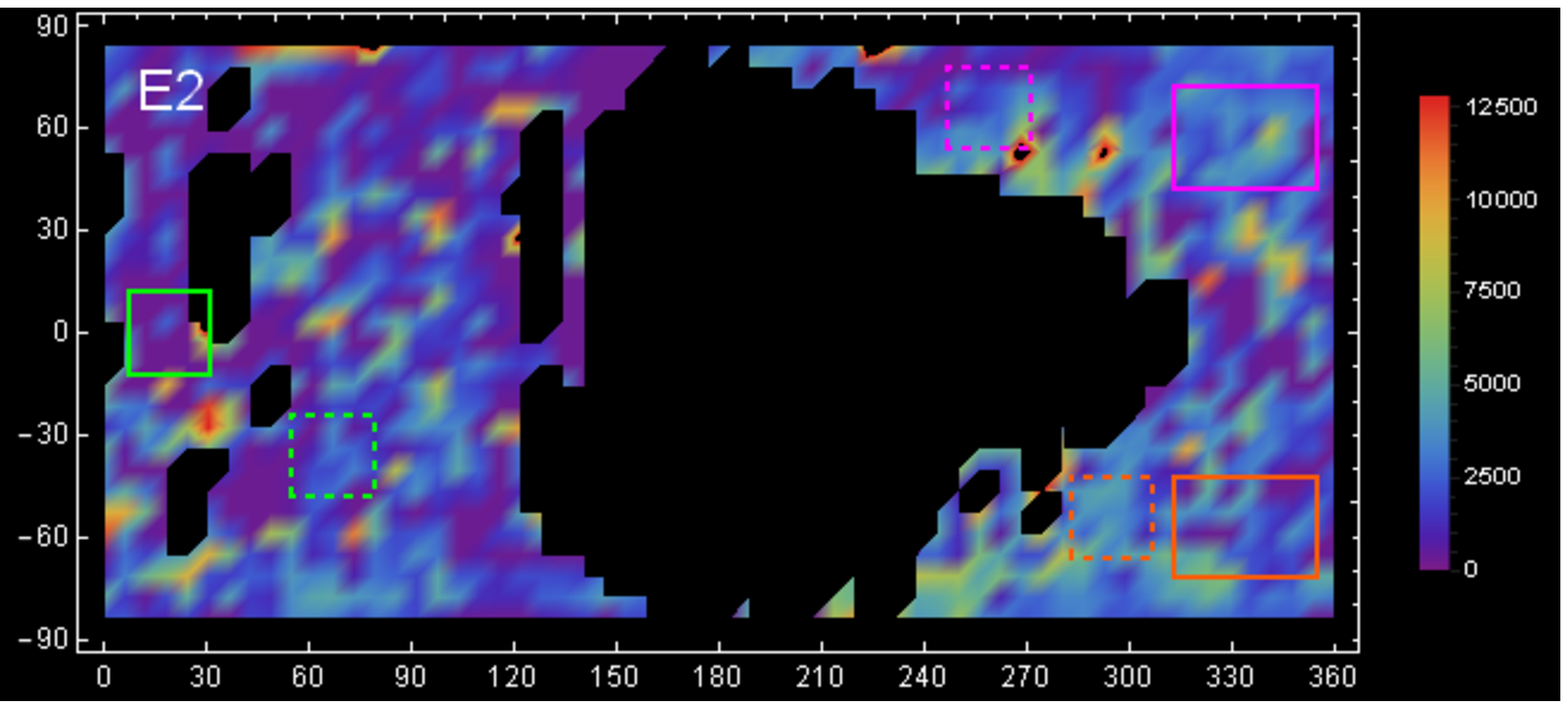}

\plottwo{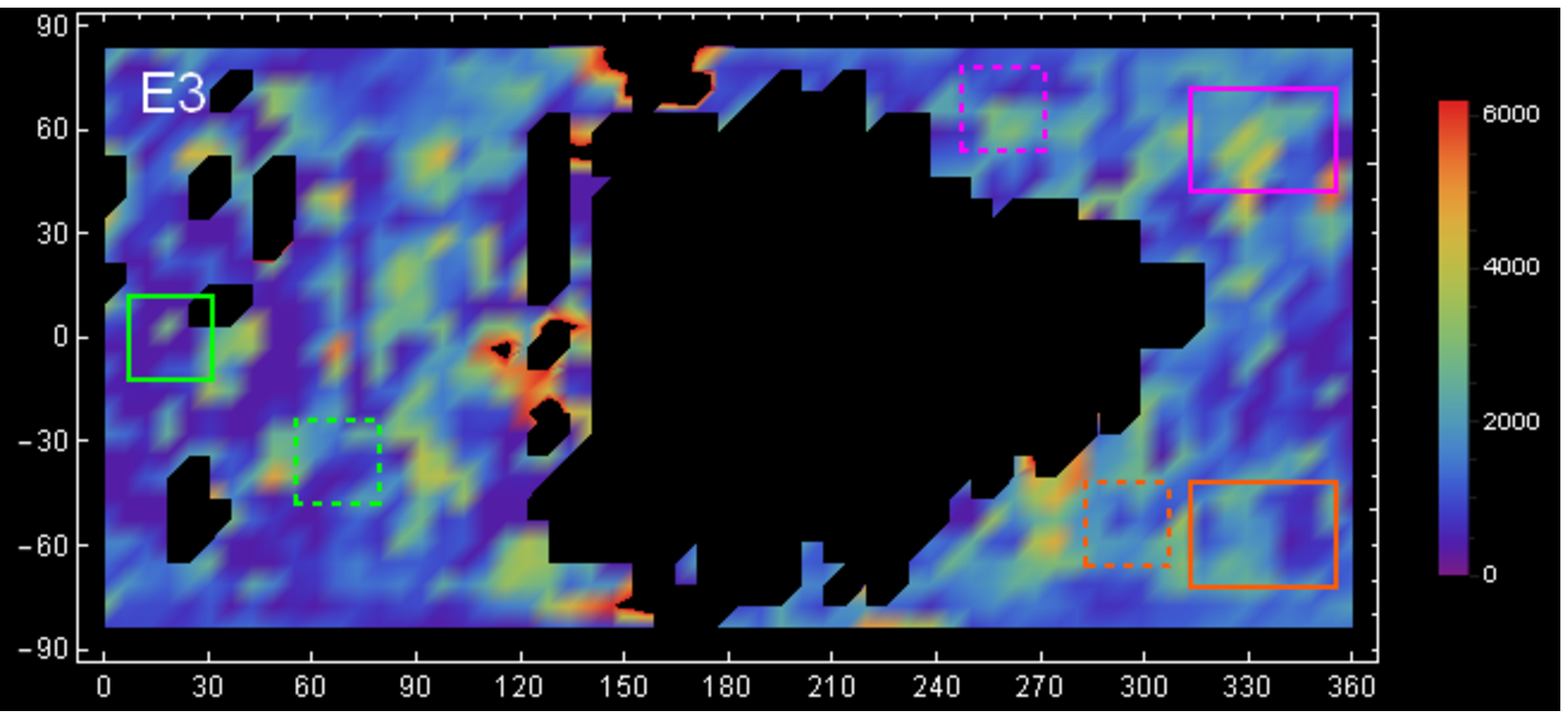}{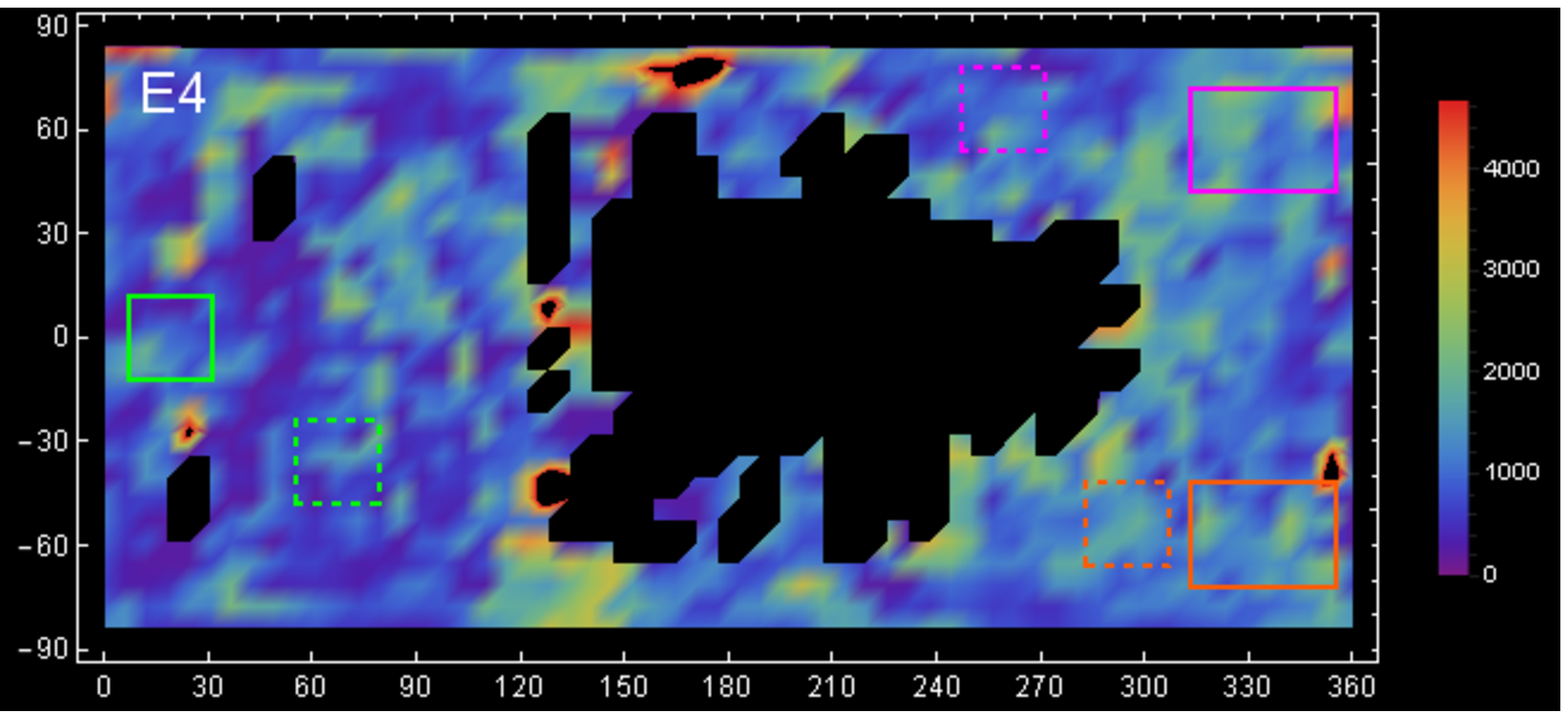}

\plottwo{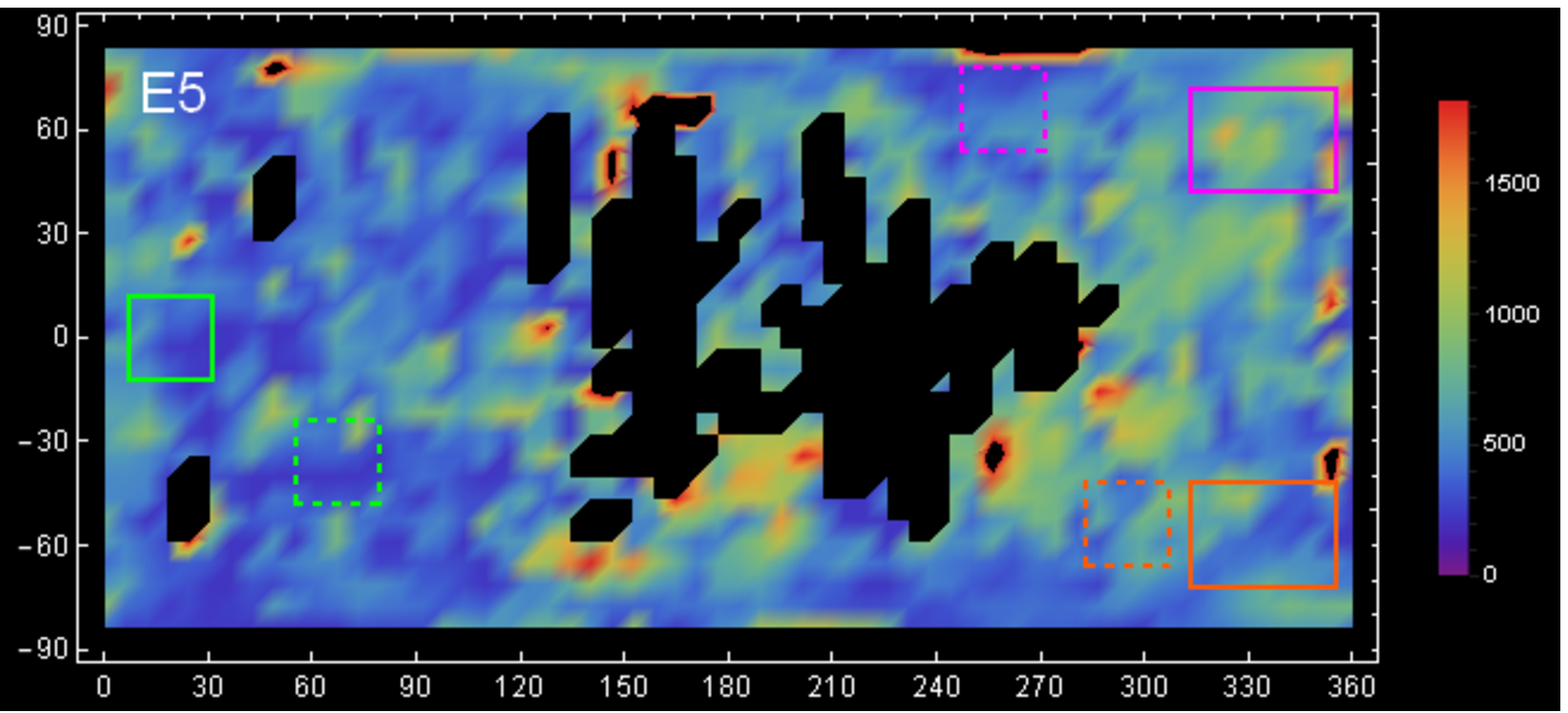}{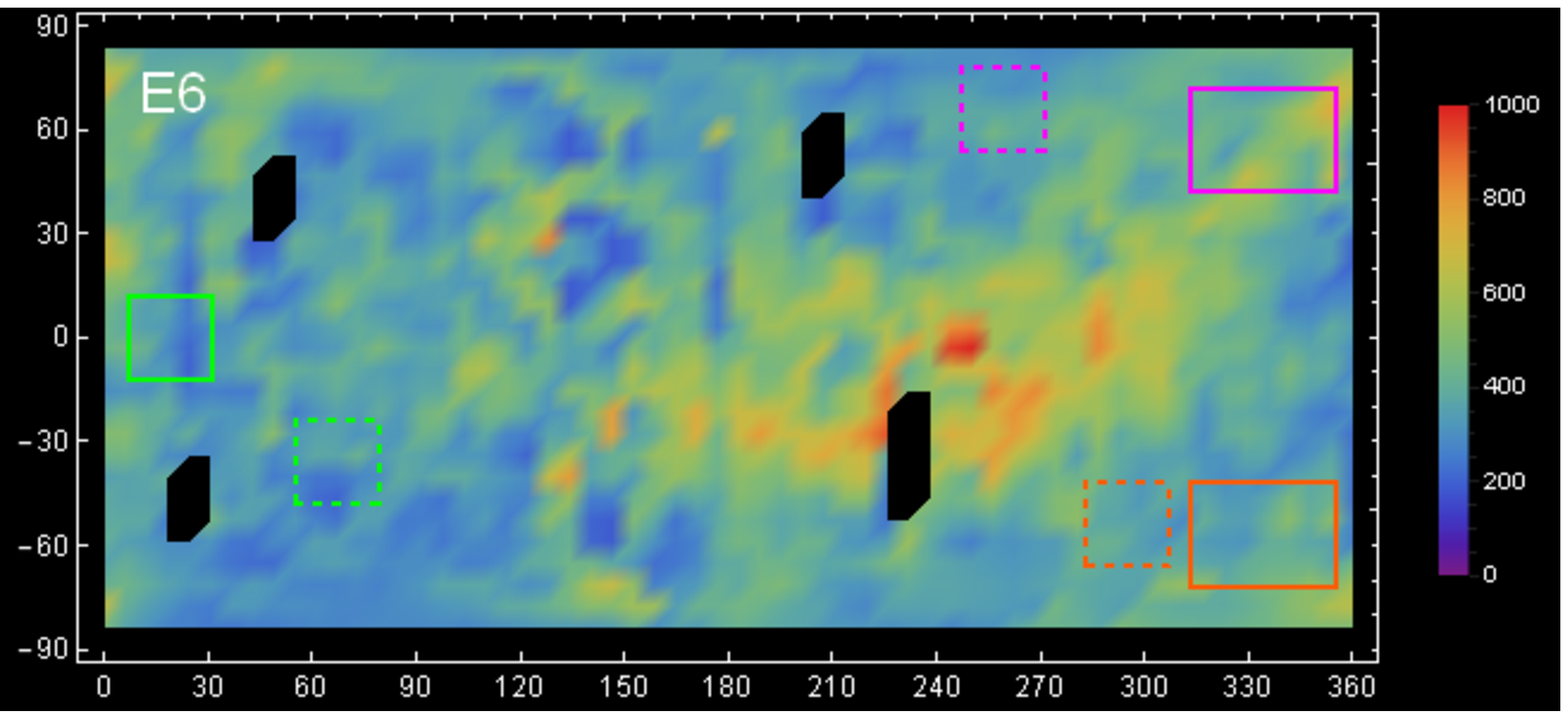}

\plottwo{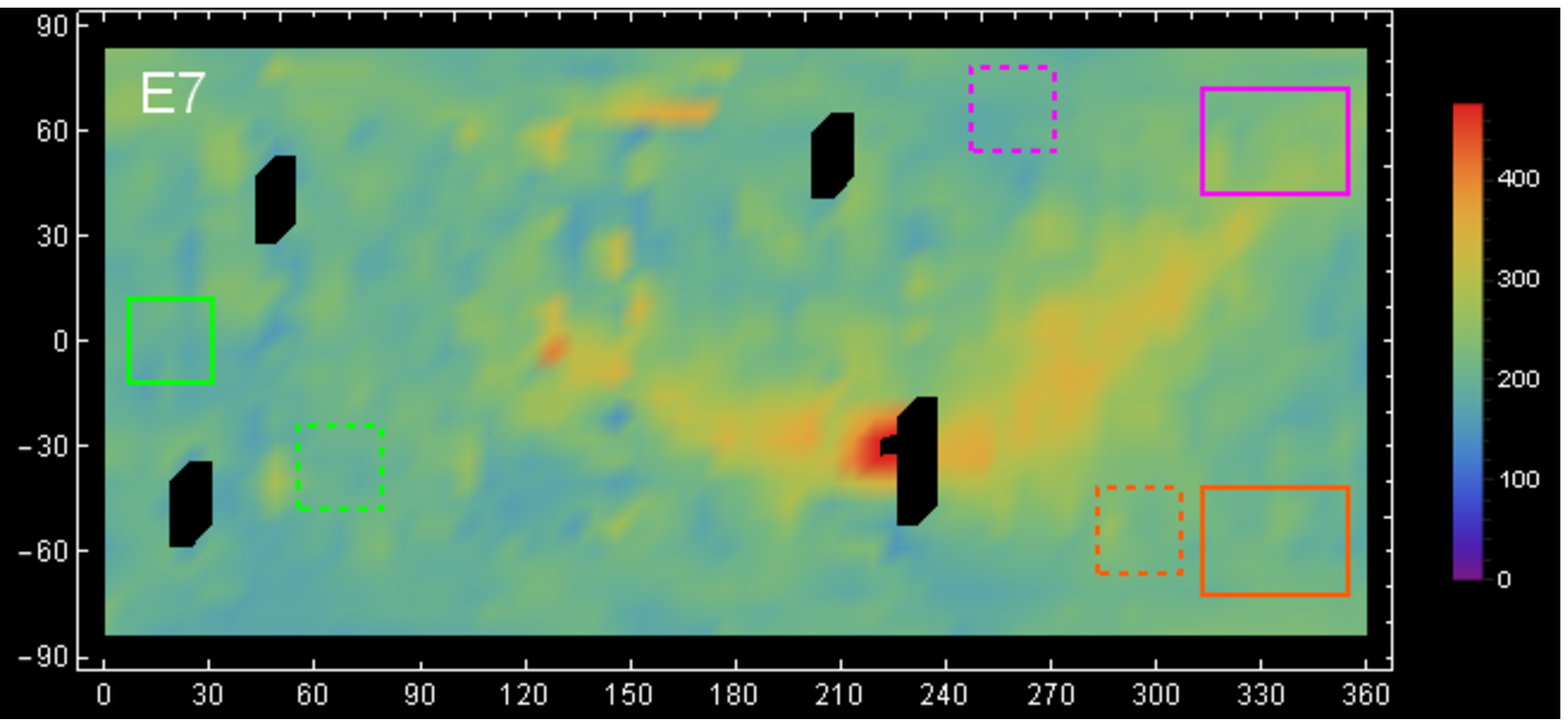}{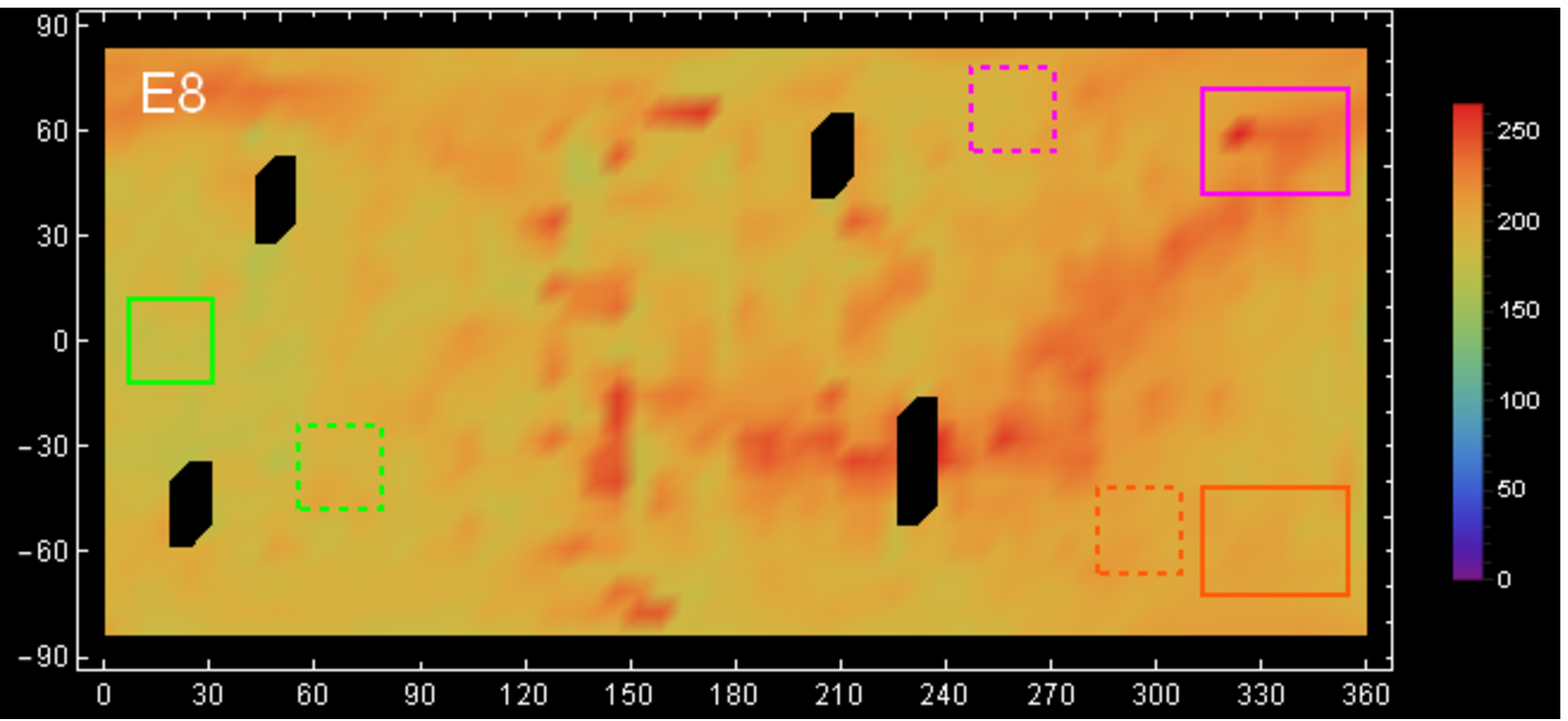}
\caption{Color-scale maps in the ecliptic coordinates of the ENA flux observed by IBEX-Lo in the eight energy steps of IBEX-Lo, based on observations taken between 2009.0 and 2012.5, repetaed after \citet{galli_etal:16a} to illustrate the dynamical scale of the signal distribution in the sky and to show the location in the sky of the macropixels used in this study. The colored rectangles mark the boundaries of the macropixels used to estimate the total density of H atoms, precisely corresponding to the macropixel boundaries used  by \citet{galli_etal:16a}: ``downwind'' (solid green), ``hole'' (broken green), ``north of Voyager 1'' (dashed magenta), ``north'' (solid magenta), ``south of Voyager 2'' (broken red), ``south'' (solid red). The horizontal axes correspond to ecliptic longitude, the vertical axes to ecliptic latitude. The large dark-blue regions in panels E1 through E4 represent the regions masked because of the presence of ISN H or a contamination.}
\label{fig:maps}
\end{figure*}

IBEX \citep{mccomas_etal:09a} is a spin-stabilized spacecraft orbiting the Earth in a high, elongated orbit, with the boresight of its two ENA detectors directed perpendicular to the spin axis. As the spacecraft is traveling around the Sun with the Earth, the direction of the spin axis is maintained within a few degree from the Sun. With this, it is possible to make yearly full-sky maps of the distribution of the ENA flux. For details of the geometry of IBEX observations see, e.g., \citet{fuselier_etal:14a, galli_etal:14a}.

\begin{deluxetable*}{ccccccccccc}
\tablecaption{Central energies and central speeds of IBEX-Lo energy steps E1--E8; the observed fluxes, corrected for the Compton-Getting effect for the downwind, Voyager~1, Voyager~2  hole, north, and south macropixels, defined in \citet{galli_etal:16a} and shown in Figure~\ref{fig:maps}; the arithmetic average flux from the north and south macropixels; and partial H densities based on the north-south averaged flux}
\tablehead{
\colhead{ } & \colhead{$E$~[keV]} & \colhead{$v$~[km/s]} & \colhead{dnwind\tablenotemark{*}} & \colhead{Voyager~1\tablenotemark{*}} & \colhead{Voyager~2\tablenotemark{*}} & \colhead{hole\tablenotemark{*}} & \colhead{north\tablenotemark{*}} & \colhead{south\tablenotemark{*}} & \colhead{average\tablenotemark{*}\tablenotemark{NS}} & \colhead{density [cm$^{-3}$]\tablenotemark{\tiny{NS}} \tablenotemark{$\dagger$}} }
\startdata
 \text{E1} & 0.015 & 53.6065 & 23330.5 & 626.111 & 389.059 & 579435. & 1997.32 & 331.615 & 1164.47 & $2.9\times 10^{-5}$ \\
 \text{E2} & 0.029 & 74.5367 & 5733.42 & 889.993 & 948.906 & 63018.1 & 3055.78 & 2395.85 & 2725.82 & $9.3\times 10^{-5}$ \\
 \text{E3} & 0.055 & 102.649 & 2217.73 & 541.242 & 1050.28 & 7160.91 & 1654.43 & 1209.29 & 1431.86 & $6.7\times 10^{-5}$ \\
 \text{E4} & 0.110 & 145.167 & 1128.43 & 561.090 & 748.263 & 2109.17 & 1455.73 & 1081.43 & 1268.58 & $8.5\times 10^{-5}$ \\
 \text{E5} & 0.209 & 200.099 & 518.338 & 275.285 & 462.745 & 712.217 & 695.534 & 381.887 & 538.711 & $4.9\times 10^{-5}$ \\
 \text{E6} & 0.439 & 290.004 & 167.356 & 148.087 & 185.536 & 261.120 & 291.225 & 200.720 & 209.007 & $3.3\times 10^{-5}$ \\
 \text{E7} & 0.872 & 408.724 & 100.418 & 61.9960 & 101.475 & 96.2042 & 137.890 & 89.1591 & 97.8571 & $2.1\times 10^{-5}$ \\
 \text{E8} & 1.821 & 590.645 & 50.0016 & 48.7235 & 57.0213 & 36.2584 & 80.4607 & 52.1472 & 54.1021 & $1.8\times 10^{-5}$ \\
\enddata
\tablenotetext{*}{observed flux in the units (cm$^2$~s~sr~keV)$^{-1}$}
\tablenotetext{\dagger}{Uncertainties for the north and south pixels estimated to be a factor of 10 for E1, E2; factor of 2 for E3; 50\% for E4; and 30\% of relative uncertainty for E5--E8, after \citet{galli_etal:16a}}
\tablenotetext{\text{\tiny{NS}}}{ over the north and south pixels}
\label{tab:data}
\end{deluxetable*}

The observation time of IBEX-Lo is divided between eight spectral channels, the so-called energy steps. They are sequentially switched during the spacecraft rotation. When the instrument is set to a given energy step, the data collected within one spacecraft rotation, which is approximately 15 seconds long, are divided into equal time blocks. As a result, the measurements for a given spacecraft orbit and energy step are collected in 60 bins in the spacecraft spin angle. Observations collected during the time interval when the spin axis is maintained fixed cover a great-circle strip in the sky with a width of $\sim 6\degr$. After six months of observations, data from individual orbits are re-binned into a full-sky map in the ecliptic coordinates. The $6\degr\times 6\degr$ pixels in these latter maps are based on the ecliptic coordinate grid and consequently they are not equi-areal globally, but equi-areal within latitudinal bands.

The energy channels of IBEX-Lo are relatively wide in the energy space, with $\Delta E/E \simeq 0.7$. When set to a given energy step $i$, the instrument is sensitive to atoms with the kinetic energies $E_i \pm 0.35 E_i$, with the center energies of the energy steps listed in Table~\ref{tab:data} after \citet{galli_etal:16a}. In the spacecraft-inertial frame, the energies of the atoms measured within a given energy step are equal within the width of the energy step. 

The maps of IBEX-Lo observations, meticulously cleaned from background contamination, were composed by \citet{galli_etal:14a, galli_etal:16a} and \citet{galli_etal:17a}. They are shown in Figure~\ref{fig:maps}. In addition to GDF, they feature the IBEX Ribbon. In this work, we only focus on the GDF component. The ISN H and Ribbon are considered as a foreground. The ISN H region was masked in the panels of Figure~\ref{fig:maps} corresponding to energy steps 1--4. The Ribbon is the most prominent in the high-energy energy steps, starting from $\sim 400$~eV, and covers a relatively small portion of the sky: with the angular radius of $\sim 75\degr$ \citep{funsten_etal:13a, funsten_etal:15a} and the width of $\sim 30\degr$ \citep{schwadron_etal:11a}, the Ribbon area is about $\pi$~steradian, i.e., 25\% of the sky area. According to present views \citep[e.g.,][]{zirnstein_etal:16a, swaczyna_etal:16a, swaczyna_etal:16b}, the emission of Ribbon ENAs originates beyond the heliopause. When projected on the sky, it is superimposed on the regular GDF emission. Most likely, the GDF covers the entire sky, including the regions occupied by the ISN signal and the Ribbon. For the purpose of this paper, we approximate the GDF flux as uniform in the sky since the uncertainties of the observed fluxes are much larger than potential spatial variabilities at the low energies, which contribute most to the GDF hydrogen density \citep{galli_etal:16a}. The GDF does show spatial features at higher energies, as discussed, e.g., \citet{schwadron_etal:14a} and \citet{schwadron_etal:18a}. These spatial patterns include, among others, enhancements by a factor of 2--3 in $\sim \pm 30\degr$ areas around the nose and/or tail region, depending on the energy. The uncertainties related to the non-uniformity of GDF on the density of the GDF population are discussed at the end of Section~\ref{sec:calculations}.

The maps shown in Figure~\ref{fig:maps} present data collected between 2009.0 and 2012.5. Following the  discussion presented in \citet{galli_etal:16a}, we consider six macro-pixels identical to those used by these authors. The names of these macro-pixels are listed in the header of Table~\ref{tab:data}. The macro-pixel boundaries are plotted in the maps shown in Figure~\ref{fig:maps}. The flux uncertainties are caused by a low signal-to-noise ratio and by various background sources that in many $6\degr$ by $6\degr$ pixels are comparable or even more intense than the GDF signal at energies below 200~eV. 

We begin the analysis from the raw flux values from the selected macro-pixels. We use the cleaned subset of observations collected during first eight 6-month full-sky imaging campaigns between 2009.0 and 2012.5. The data selection and cleaning procedure is presented in detail in \citet{galli_etal:16a}, but unlike in this latter paper, we start with the filtered and the ubiquitous background subtracted, but otherwise uncorrected data in the spacecraft frame. The magnitudes of the ubiquitous background are listed in \citet{galli_etal:15a}. The measured flux values corrected for the Compton-Getting effect, the central energies of the IBEX-Lo energy steps, and the corresponding central speeds of H atoms for these steps are listed in Table~\ref{tab:data}. The fluxes in the table are given in the absolute units (cm$^2$~s~sr~keV)$^{-1}$. These quantities are differential spectral flux values, averaged over the area of the macro-pixels and over the energy bands of IBEX-Lo energy channels (``energy steps''), and therefore can be used directly to calculate the absolute density of the low-energy ENA population at 1~au, as described in Section \ref{sec:calculations}. 

\citet{fuselier_etal:14a} and \citet{galli_etal:14a} pointed out that most of the IBEX-Lo data for low-energy steps have a very significant local co-moving foreground component of unknown nature. The telltale signature is the behavior of the signal observed in a given region in the sky when observed in the ram ant anti-ram viewing geometry, i.e., half year apart, when the spacecraft with the Earth are moving in the opposite directions in the travel around the Sun. Assuming a relatively slow  variation in the GDF, the Compton-Getting corrected fluxes for the same pixels in the ram and anti-ram viewing should be close to each other, unless there is a local foreground component, co-moving with the spacecraft. In that case, the measured Compton-Getting corrected fluxes will be different in the ram and anti-ram viewing because the hypothetical co-moving component must not be corrected for the Compton-Getting effect.

\begin{figure}
\plotone{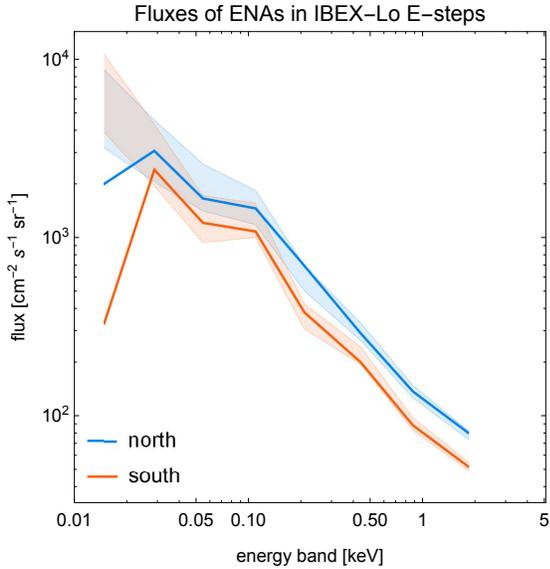}
\caption{Spectral fluxes of GDF H for the centers of all energy steps of IBEX-Lo for the north and south macro pixels (blue and orange lines). The time-averaged spectral flux, obtained from Compton-Getting corrected fluxes for individual half-year maps, are expected to fit into envelopes formed by time-averaged uncorrected fluxes, with the Compton-Getting correction applied assuming either the entirely ram, or entirely anti-ram viewing. }
\label{fig:NoForegroundNS}
\end{figure}

This reasoning enables formulating a sanity test for the adopted data. In the absence of a co-moving foreground, the time-averaged flux in individual energy bins, originally uncorrected for the Compton-Getting effect and subsequently corrected assuming either ram or anti-ram viewing, will produce two spectra that will provide an envelope for the actually-observed time-averaged flux, when the corrections for the observations were performed for individual semi-annual maps with the appropriate viewing geometry taken into account. An excursion of the averaged Compton-Getting corrected spectra outside the envelope flags the presence of a co-moving component.

We have performed this sanity test for our selected macro-pixels. We have taken the measured uncorrected fluxes averaged over the time interval of the observations and applied the Compton-Getting correction as described by \citet{mccomas_etal:10c} in Appendix~A assuming that all the data from a given macro-pixel were collected either in the ram or the anti-ram viewing geometry. The two envelope spectra for each macro-pixel were subsequently compared with the corresponding spectrum averaged over the observation interval from 2009.0 to 2012.5, where the average values for the individual energy steps were computed from Compton-Getting corrected individual pixel observations.  We found that only the north and south macro-pixels pass this sanity test. In all remaining ones, the three or four lowest-energy steps bore signatures of a co-moving foreground. 

This test for these macro-pixels is presented in Figure~\ref{fig:NoForegroundNS}. Clearly, the test is passed except for energy step 1. In all other macro-pixels, the three or four lowest-energy steps bore signatures of a co-moving foreground. It is not surprising that out of the macro-pixels originally considered by \citet{galli_etal:16a} these two ones turn out to be the best: this is because they are in high latitudes and therefore, owing to the observation geometry, they have the longest time coverage and consequently the best observation statistics with an equal amount of ram and anti-ram observations from the macro-pixels considered by \citet{galli_etal:16a}. Therefore, for the calculation of GDF H density presented in the further part of the paper, we adopted the north and south macro-pixels, which feature a non-zero flux. A non-vanishing GDF H ENA flux for these energies is predicted by certain models heliospheric models \citep{zirnstein_etal:18c}. Since, on the other hand, the contribution from the co-moving foreground to these energies is not well understood, we attribute to the two lowest-energy steps and the energy band between 0 and the lower limit of E1 a large uncertainty of a factor of 10. 

\section{Calculations}
\label{sec:calculations}

In this section, we calculate partial densities of neutral H atoms from GDF ENA at 1~au, observed in individual energy steps of IBEX-Lo. With them, we estimate the total density of this neutral H component. 

In general, calculation of the total density of a particle population in an inertial reference frame in a given location in space is done by integration of the distribution function $f(v, \theta, \phi)$ of this population in this location over the entire velocity space, as defined in a spherical coordinate system in the following equation:
\begin{equation}
n = \int\limits_0^{2 \pi}\int \limits_{-\pi/2}^{\pi/2} \int \limits_{0}^{\infty} f(v, \theta, \phi) v^2 \,\cos \theta \, d \theta d\phi.
\label{eq:densityDefinition}
\end{equation}
The result does depend on the choice of the inertial reference frame. However, we do not have the distribution function but only maps as a function of the direction in the sky of the differential flux in several energy bands. Therefore, we are only able to make an approximate assessment of the density at 1~au. The differential flux $F_E(\vec{l})$ in a given map pixel with the center given by a direction $\vec{l}$, defined by the (longitude, latitude)~=~$(\phi, \theta)$ for energy step $E$ is defined as
\begin{equation}
F_E(\vec{l}) = \int \limits_{\phi_1}^{\phi_2} \int \limits_{\theta_1}^{\theta_2} \int\limits_{v_{E,1}}^{v_{E,2}} v\, f(v, \vec{l}) v^2 dv\, \sin \theta d \theta d \phi
\label{eq:partialFlux}
\end{equation}
where the integration goes over the speed $v$ within the boundaries $(v_{E,1}, v_{E, 2})$ determined by the energy boundaries of this given energy step $E$ and over the spherical coordinates of the pixel boundaries. In this formula, $f(v, \vec{l})$ is the magnitude of the distribution function for speed $v$ and direction $\vec{l}$ in space. This definition does not depend on the functional form of the distribution function. The partial density $n_E(\vec{l})$ for a pixel centered at $\vec{l}$ (i.e., the contribution to the total density in a given location in space from atoms observed within a given pixel) is given by:
\begin{equation}
n_E(\vec{l}) =\int \limits_{\phi_1}^{\phi_2} \int \limits_{\theta_1}^{\theta_2} \int\limits_{v_{E,1}}^{v_{E,2}} f(v,\vec{l}) v^2 dv \sin \theta\, d \theta d \phi.
\label{eq:partialDens}
\end{equation}

Approximately, within a given energy step with the mean energy $E$, the flux can be approximated by 
\begin{equation}
F_{\mathrm{appr, E}} = \langle v_{\mathrm{E}} \rangle n_{\mathrm{E}},
\label{eq:approxEStepFlux}
\end{equation}
where $\langle v_{\mathrm{E}}\rangle$ is the mean atom speed for the given energy step, and $n_E$ is the partial density. Hence, given the measured flux value for a given energy step $F_{\mathrm{data,E}}$, one calculates the partial density for this pixel as:
\begin{equation}
n_{\mathrm{E}}(\vec{l}) = F_{\mathrm{data,E}}(\vec{l})\, \Delta E \,\Delta \Omega (\vec{l})/\langle v_{\mathrm{E}}\rangle,
\label{eq:partialDensMeasured1}
\end{equation}
where $\Delta E$ is the width of the energy step and $\Delta \Omega$ is the pixel area.
Summing the partial densities over the entire map, we obtain the partial density for the energy step E. Subsequently, summing over the partial densities for individual energy steps, we obtain the total density of ENAs:
\begin{equation}
n = \sum \limits_{E = E1}^{E8} \sum\limits_{\vec{l}} n_E(\vec{l}) .
\label{eq:partialDensMeasired2}
\end{equation}

In the calculations, we first estimated the globally-averaged GDF differential flux. We did this by averaging the fluxes observed in each energy step over the six macro-pixels shown in Figure~\ref{fig:maps}. The magnitudes of the flux for these macro-pixels for each energy step are listed in Table~\ref{tab:data}. The mean flux for a given energy step $E$ was calculated as
\begin{equation}
\langle F_E \rangle = \left(\sum \limits_i \langle F_{E,i}\rangle \Delta \Omega_{E,i}\right)/\sum\limits_i \Delta \Omega_{E,i}.
\label{eq:partialFlux2}
\end{equation}
In this formula, $\langle F_{E,i}\rangle$ is the partial flux in energy step $E$ averaged over the macro-pixel area $i$, and $\Delta \Omega_{E,i}$ is the effective area of macro-pixel $i$ for the energy step $E$. These effective areas for a given macro-pixel may vary  between energy steps because for some of the pixels included in a given macro-pixel the actual value of the observed flux may be unavailable because of high background or other reasons, as discussed in detail by \citet{galli_etal:14a, galli_etal:16a, galli_etal:17a}. The pixel-averaged partial fluxes and mean partial flux values from the Compton-Getting averaged fluxes from the north and south macro-pixels are listed in Table~\ref{tab:data}.

Having evaluated Equation~(\ref{eq:partialFlux}) for maps collected in all eight energy steps we obtained a list of differential mean fluxes that we adopted as representative for the entire sky, expressed in the units of atom$(\text{s cm}^2\, \text{eV sr})^{-1}$. With them, we could calculate the partial densities $\langle n_E \rangle$ from the entire sky for the individual energy steps. 
\begin{equation}
\langle n_E \rangle = \langle F_E \rangle \Delta E (4 \pi)/\langle v_E \rangle,
\label{eq:partialDensities}
\end{equation}
where $\Delta E$ is the width of the energy step in energy space, $(4 \pi)$ is the full-sky volume angle, and $\langle v_E\rangle$ is the speed corresponding to the central energy $E$ of a given energy step. The partial densities thus obtained are shown in Figure~\ref{fig:partialDensities} and listed in Table~\ref{tab:data}.

Since IBEX observations do not cover very low energies but we need to integrate over the entire velocity space, we assumed that the density of H at 0 speed is equal to 0 and we approximate the contribution of the lowest-energy H atoms as half of the contribution from the lowest IBEX energy step. Thus, the total density $n$ of GDF H atoms at 1~au from the Sun is obtained as
\begin{equation}
n = \frac{1}{2} \langle n_{\mathrm{E1}}\rangle + \sum \limits_{E=\text{E1}}^{\text{E8}} \langle n_E \rangle
\label{eq:totalDensity}
\end{equation}
The first term in Equation~(\ref{eq:totalDensity}) is an educated speculation based on the assumption that at 0 velocity all H atoms at 1~au are expected to be eliminated by ionization. However, the presence or absence of the rollover of the GDF flux at low energies is not certain \citep{galli_etal:16a, zirnstein_etal:18c}. Anyway, the contribution to the total density of the partial density from energy step 1 is $\sim 7$\%, so the first speculative term in Equation~(\ref{eq:totalDensity}) is only 3.5\% of the reported value, which is small compared with the large overall uncertainty of the result (a factor of 10, see the Conclusions section). 

We cut off the contributions from energies higher than the energy of the highest energy step of IBEX-Lo. We believe this adds little to the uncertainty of the result, because the partial flux is a rapidly decreasing function of energy and the contribution from the highest energy step to the total density is only $\sim 4$\%. 

The likely presence of the co-moving foregroud in the lowest energy step does not invalidate our result since anyway the contribution of this step to the overall density as we obtained is only $\sim 7$\%. The contribution of the four lowest-energy bins to the overall density is $\sim 50$\%, as illustrated in the lower panel of Figure~\ref{fig:partialDensities}.

The IBEX ENA maps are obtained from observations of atoms detected when they are close to the perihelion in their orbits. This is due to the specific IBEX observation geometry: the detectors are looking perpendicular to the spacecraft--Sun lines. We assume that a semi-annual map for a given energy step is representative for the global full-sky flux distribution of ENAs in all locations along the Earth's orbit in this energy band. In other words, we assume that if IBEX had the capability to observe the ENAs from the entire sky while being in a given location in space, the maps obtained at various locations along the Earth's orbit would be identical to each other and to the maps that have been obtained from the actual observations. 

This assumption is only an approximation. In reality, an instrument with an instantaneous full-sky mapping capability would most likely find that the flux of the atoms that arrive from a certain angular range from the Sun is more suppressed by ionization than the flux of the atoms arriving from directions that never bring them close to the Sun. This suppression radius is most likely a decreasing function of atom energy. However, this suppression is only able to reduce the total density by a factor of 2 (it concerns at most only half of the sky); a reduction by a factor of 2 would be obtained if atoms from the entire sunward hemisphere were fully suppressed, which is not likely to be the case. Given all other uncertainties, we believe this simplification is justifiable. 

Another uncertainty comes from the non-uniform distribution of GDF in the sky. Specifically, enhancements by a factor of 2--3 were observed in some energy steps from regions of $\sim \pm 30\degr$ around the upwind and downwind directions. It is not clear if the upwind enhancement extends to the lowest energies viewed by IBEX, which contribute the most to the GDF population density, because the upwind region is covered by the ISN H background (cf. Figure~\ref{fig:maps}). In the downwind region, an enhancement in the lowermost energy bins is not visible, but these observations are affected the most by the ubiquitous background. Various regions in the sky feature various spectra, as discussed by \citet{desai_etal:16a} who point out that spectral indices of ENAs in the energy range observed by IBEX in the Ribbon differ from those in the upwind and downwind directions and for the GDF. Hence, there is no single feature in the ENA sky that has been shown to consistently persist in all energy steps. To assess an uncertainty resulting from neglecting the large-scale persistent features, like the upwind and downwind enhancements, we assumed they occupy regions within $30\degr$ around the upwind and downwind directions, i.e., occupying $\frac{1}{4}$ of the sky area each, and featuring an enhancement by a factor of 2 in all energy steps over the mean value of GDF. Then, the density would be increased by half (a factor of two enhancement from half of the sky). This is much less than our global uncertainty. 

\section{Results}
\label{sec:results}
\begin{figure}
\plotone{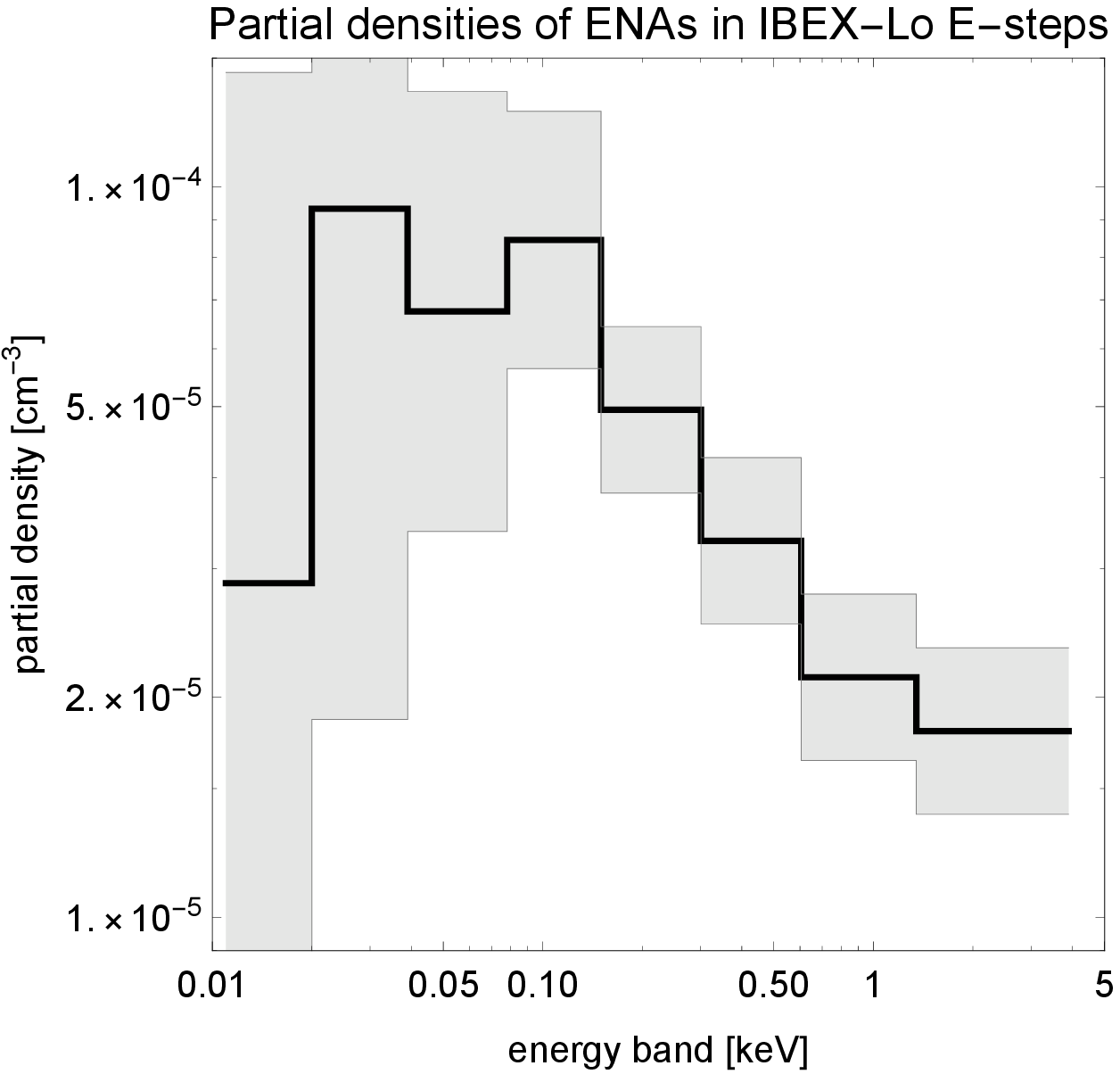}

\plotone{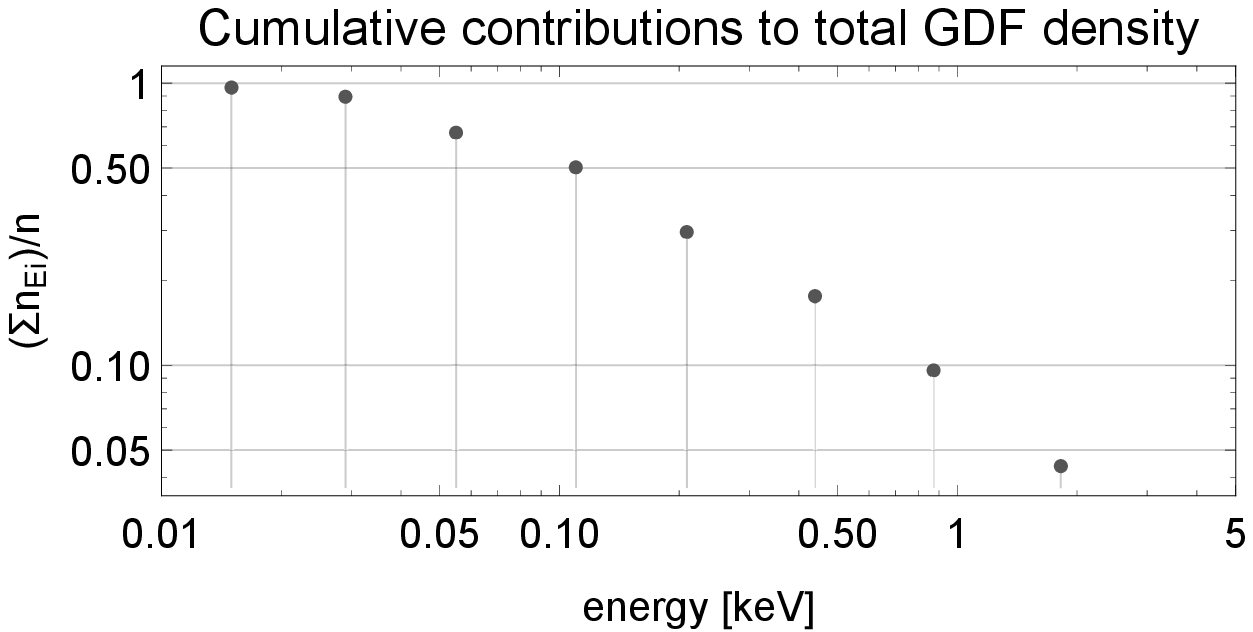}
\caption{Partial densities of GDF H calculated for all energy steps of IBEX-Lo, calculated from averaging over the north and south macro-pixels (upper panel). The widths of the horizontal line sections  correspond to the distances in energy space between the centers of the subsequent energy steps of IBEX-Lo. The uncertainties of the partial densities, marked with the gray shading, correspond to the uncertainties listed in Table~\ref{tab:data}. Cumulative contributions of the partial densities to the total ($\sum \limits_{\texttt{E8}}^{\texttt{E1}} n_{\texttt{Ei}}/n $) are shown in the lower panel.}
\label{fig:partialDensities}
\end{figure}
The results of the calculations of the partial densities for energy steps of IBEX-Lo are listed in Table~\ref{tab:data}. A plot of these densities is shown in Figure~\ref{fig:partialDensities}. The total density of neutral H, obtained from the partial densities using Equation~\ref{eq:totalDensity}, is equal to $4.1\cdot 10^{-4}$~cm$^{-3}$. The mean energy of these atoms is calculated as
\begin{equation}
\langle E\rangle = \left(\sum n_i E_i\right)/\sum n_i,
\label{eq:meanEnergy}
\end{equation}
where $i$ corresponds to the numbers of IBEX energy steps; $n_i$ are the partial densities, and $E_i$ are the central energies of the energy steps. The mean energy thus calculated is equal to 0.22~keV and the mean atom speed of $\sim 200$~km~s$^{-1}$, obtained from the formula
\begin{equation}
\langle v\rangle = \left(\sum n_i v_i\right)/\sum n_i
\label{eq:meanSpeed}
\end{equation}

\begin{figure}
\plotone{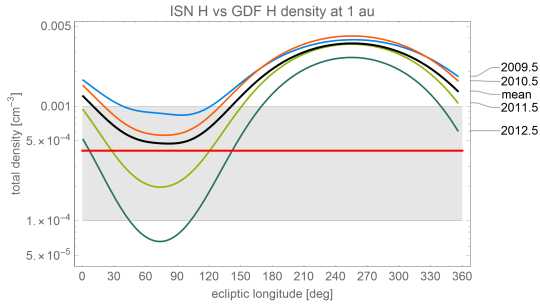}
\caption{Comparison of the density of GDF H, observed by IBEX-Lo at $4.1 \cdot 10^{-4}$~cm$^{-3}$, represented by the red horizontal line, with the distribution of ISN H density at 1~au along the ecliptic plane, averaged in time over the GDF observation interval 2009.0--2012.5 (black thick line) and the ISN H densities for selected epochs within this interval (colored lines). The gray-shaded region marks the uncertainty range. The ISN H density along the ecliptic plane averaged over ecliptic longitudes and over the time interval 2009.0--2012.5 is equal to $2.3 \cdot 10^{-3}$~cm$^{-3}$.}
\label{fig:ISNComparison}
\end{figure}

Figure~\ref{fig:ISNComparison} presents a comparison between the density of the GDF population and the densities of ISN H (a sum of the primary and secondary population), averaged over the observation interval 2009.0--2012.5, with the density distributions along the Earth orbit calculated for each half of the year (2009.0, 2009.5, 2010.0,...,2012.0) using the state of the art model of ISN H in the heliosphere (Sok{\'o}{\l} et al. 2018, in preparation) with the realistic, observation based model of the solar resonant radiation pressure \citep{kowalska-leszczynska_etal:18a} and of the ionization rates \citep{sokol_etal:19a}. In addition to the interval-averaged ISN H density, also the ISN H densities for the epochs of halves of years 2009 through 2012 are shown.  

Clearly and not surprisingly, the GDF H density at 1~au is lower than the ISN H density. However, close to the downwind region (see a band of ecliptic longitudes centered at $\sim 75 \degr$) these two densities become comparable ($n_{\texttt{GDF H}}/n_{\texttt{ISN H}}(\mathrm{downwind}) \simeq 0.9$) and may be the GDF H density near the downwind region exceeds that of ISN H at certain epochs. Even at the upwind longitudes, $n_{\texttt{GDF H}}/n_{\texttt{ISN H}}(\mathrm{upwind}) \simeq 0.1$. 

\section{Discussion}
\label{sec:discussion}
Our assessment of the density of the low-energy ENA population is accurate to no better than an order of magnitude. A hypothetical non-uniformity of the GDF distribution in the sky at lowest energies would imply variations in the GDF H density along the Earth orbit, as mentioned in Section~\ref{sec:calculations}. To reach the selected location at 1~au, the neutral atoms must pass close to the Sun, which makes them susceptible to enhanced ionization losses. If in a given location at 1~au the portion of the sky with an elevated flux level is partially blocked by the Sun, then the total density in this location is reduced in comparison with other locations in the ecliptic. Specifically, if flux enhancements in the lowest energy steps exist in the upwind and downwind portions of the inner heliosheath, then the GDF H density observed at 1~au in the ecliptic plane close to crosswind locations is expected to be larger than in the upwind and downwind regions at the Earth orbit, because in the latter locations either the downwind or the upwind flux enhancements are partially blocked by the Sun. This effects should be detectable provided that ENAs with the energies below 200~eV can be observed from the entire sky.

A more detailed studies of the variation of the density of GDF population of atoms as a function of the location along the ecliptic plane and at different distances from the Sun require simulations with a realistic assumption on the distribution function of these atoms at their source and detailed assessment of the ionization losses, effects of radiation pressure etc. along the trajectories of these atoms from the source to a given location. Since half of these atoms have a relatively low velocity at 1~au, they are sensitive to the solar radiation pressure and its variation with the solar activity and with the atom radial speed due to the Doppler effect and the self-reversed shape of the solar Lyman-$\alpha$ line, responsible for the radiation pressure effect. These studies are left for a future work. However, notwithstanding all the uncertainties, the presence of a population of H atoms from GDF at 1~au is a reality since the flux of these atoms has been directly measured in situ by IBEX-Lo. Here we wish to point out some of the potential consequences of the existence of this population for the studies of the heliosphere.

Half of the density of this population of H atoms have energies within the four lowest energy steps of IBEX. This implies that their energies are only a factor 2 to 3 larger than the energies of ISN H atoms at 1~au \citep[cf. typical ISN H energies observed by IBEX in ][]{kowalska-leszczynska_etal:18b}. The width of the solar Lyman-$\alpha$ line corresponds to $\pm 200$~km~s$^{-1}$ radial speed \citep{lemaire_etal:15a, kowalska-leszczynska_etal:18a}. Consequently, these atoms interact with the solar Lyman-$\alpha$ radiation and contribute to the heliospheric backscatter glow. Especially in the downwind hemisphere, this contribution may form an important local foreground the resonant backscatter glow of ISN H that, to our knowledge, has never been taken into account in the analyzes of the heliospheric glow. 

The slow GDF atoms are ionized by charge exchange and photoionization, forming pickup ions, as pointed out by \citet{schwadron_mccomas:10a}. In the downwind hemisphere within a few au from the Sun, the contribution from these atoms to the PUI flux may be comparable to the contribution from ISN H. Closer to the Sun, this contribution is likely even larger than at 1 au because due to the relatively large speed ($\sim 160$~km~s$^{-1}$), these atoms are expected to be able to penetrate close to the Sun. Eventually, they will be ionized and they will feed the PUI flux. However, the radial profile of the injection rate of the PUIs from the GDF H atoms will be little dependent on the ecliptic longitude (and also, likely, on the ecliptic latitude), and different from the radial profiles of the injection rate of PUIs created from ISN H atoms. It can be expected that the injection rate will be large close to the Sun, in the region where an inner source of PUIs has been discovered \citep{gloeckler_geiss:98a, gloeckler_etal:00a}. 

Recently, \citet{hollick_etal:18b} reported an excess in the number of PUI wave excitation events observed by the Voyager spacecraft shortly after launch, when they were located in the downwind hemisphere close to 1~au, in comparison with the numbers of events expected if only ISN H and ISN He atoms were the source of these PUIs. These authors suggested that this may be a signature of the presence of an inner source of PUIs of unspecified nature. We speculate that this inner source may be the population of GDF H atoms observed by IBEX.

Another aspect of the PUIs created from these atoms is that due to their large speed relative to the Sun at ionization, their injection speeds to the solar wind will be very different from 0 in the solar wind frame, unlike the newly-injected PUIs originating from ISN atoms. This will likely result in a limited-energy suprathermal tail in the distribution function of the PUIs originating from GDF ENA, expected to form already close to the Sun, without any acceleration processes operating on the PUIs themselves.

\section{Summary and conclusions}
\label{sec:conclusions}
We have analyzed full-sky-averaged maps of H ENA observed by IBEX-Lo at 1~au from the Sun between 2009.0 and 2012.5 and we have discovered that these atoms form a population with the density that may exceed the density of ISN H atoms at 1~au in a large portion of ecliptic longitudes. The absolute density of this population is $\sim 4\cdot 10^{-4}$~cm$^{-3}$, which is comparable to the density of ISN H in the ecliptic plane, averaged over ecliptic longitude and the observation interval 2009.0--2012.5. The uncertainty of this density is a factor of 10. Approximately half of the density of this population of neutral hydrogen at 1~au have energies less than $\sim 55$~eV, i.e., comparable to the energies of ISN H atoms. However, the spatial distribution of these atoms is very different from the distribution of ISN H at 1~au, and the mean energy is much larger and equals to $\sim 170$~eV. 

This result is model independent, as it was obtained directly from observations, with an accuracy of an order of magnitude. 

This GDF population of H atoms has been up to now neglected in the studies of PUIs and heliospheric backscatter glow, but it may have an important contribution to these phenomena. In particular, it may be responsible for the inner source of PUIs and for an excess in the intensity of the heliospheric backscatter glow in the downwind hemisphere. It may also be responsible for forming suprathermal tails in the PUI distribution function within a few au from the Sun and for the excess in the PUI production rate, reported based on analysis of magnetic waves related to the pickup of newly-ionized H atoms. A further assessment of the behavior of the GDF population as a function of solar distance, and consequently of the contribution to observations of PUIs, helioglow, and other effects observed away from 1~au requires careful kinetic modeling and will be a subject of future studies. 

\acknowledgments
{\emph{Acknowledgments}}. We are obliged to Marzena A. Kubiak for the calculation of ISN H density and to Stephen Fuselier for pointing out the potential importance of the co-moving foreground in the lowest-energy portion of the ENA spectrum. This study was supported by Polish National Science Center grant 2015-18-M-ST9-00036.

\bibliographystyle{aasjournal}
\bibliography{iplbib}

\end{document}